\documentclass[11pt,a4paper,oneside]{article}

\usepackage{geometry}
\usepackage{graphicx}
\usepackage{epstopdf, epsfig}
\usepackage[position=t,caption=false]{subfig}
\usepackage{amssymb}
\usepackage{float}

\usepackage{xcolor}
\usepackage{amsmath}
\renewcommand{\vec}[1]{{\mbox{\boldmath $ #1 $}}}

\newcommand{\dd}{{\rm d}}
\newcommand{\DD}{{\rm D}}

\definecolor{LightGray}{rgb}{0.7, 0.7, 0.7}

\def\Rey{\mbox{Re}}   
\def\St{\mbox{St}}    
\usepackage{natbib}
\graphicspath{{figures/}}

\newcounter{daggerfootnote}
\newcommand*{\daggerfootnote}[1]{%
    \setcounter{daggerfootnote}{\value{footnote}}%
    \renewcommand*{\thefootnote}{\fnsymbol{footnote}}%
    \footnote[2]{#1}%
    \setcounter{footnote}{\value{daggerfootnote}}%
    \renewcommand*{\thefootnote}{\arabic{footnote}}%
    }

\begin{document}

\begin{center}
\LARGE\textbf{Oscillatory switching centrifugation: \\ dynamics of a particle in a pulsating vortex}

\hspace{0.2cm}

{\large Francesco Roman\`o\daggerfootnote{e-mail: \texttt{frromano@umich.edu}}}

{\small Department of Biomedical Engineering, University of Michigan, \\ \vspace*{-0.3cm} 2123 Carl A.\ Gerstacker Building, 2200 Bonisteel Boulevard, Ann Arbor, MI 48109-2099, USA}
\end{center}

\begin{abstract}
The dynamics of a small rigid spherical particle in an unbounded pulsating vortex is considered, keeping constant the particle Stokes number $\St$ and varying the particle-to-fluid density ratio $\varrho$ and the pulsation frequency of the vortex $\omega$. We show that the asymptotic dynamics of a particle of given $\St$ and $\varrho$ can be controlled by varying $\omega$, turning the vortex core either into an attractor or a repellor. The creation of non-trivial particle limit cycles characterizes the boundaries between centrifugal and centripetal regions in parameter space. The discovered phenomenon is termed oscillatory switching centrifugation and its implications for particle demixing processes, biological protocols, {\color{black} lab-on-a-chip devices and} dynamical system theory are discussed in the end.
\end{abstract}

\textbf{keywords}: Particle/fluid flow, Nonlinear Dynamical Systems, Rotating flow

\section{Introduction}\label{sec:Intro}
The trajectory of a small rigid spherical particle immersed in an incompressible fluid flow deviates from fluid pathlines owing to its finite-size \citep{Babiano00}, particle-to-fluid density mismatch \citep{Lasheras94} and interaction with boundaries such as walls or free surfaces \citep{Brenner61}. The forces exerted by the fluid on the particle are responsible of these deviations and may lead to attraction (repulsion) of the particle towards (away from) one of the stable (unstable) manifolds of the dynamical system which governs the particle motion. These mechanisms are at the heart of particle segregation and accumulation phenomena leading to sedimentation \citep{Kynch52} and centrifugation \citep{Lasheras94}. More recently the attention of many researchers has been paid to the occurrence of subtler particulate coherent structures relying on the Lagrangian topology of the flow. The mixing of particle is promoted within regions dominated by chaotic advection, whereas either weakly chaotic areas (characterized by small Lyapunov exponent) or regular regions mark the particles settling patterns. These Lagrangian coherent structures tend, therefore, to promote accumulation structures which resemble the linearly mixing regions of the flow, occurring in form of doubly connected non-convex manifolds (Kolmogorov--Arnol'd--Moser tori) or simply connected convex manifolds (invariant spheroids). The mechanism underlining the particle coherent structures relies on the transferring of the particles from chaotic to regular or weakly chaotic regions of the flow solely due to particle finite-size \citep[neutrally-buoyant particles, see][]{Babiano00}, inertia \citep{Sapsis09} or interaction with the boundaries \citep{RomanoPRFluids}. For dilute suspensions, the demixing and accumulation of particles is typically explained as a single-particle phenomenon, on the understanding that the suspension keeps being locally dilute and particle--particle interactions are negligible. This is also the spirit of our investigation, in which we study the dynamics of a single particle, expecting that this can be representative of the dynamics of a particle suspension when particle--particle interactions are negligible. 

Among the first ones who investigated the motion of particles in rotating flows, G. I. Taylor \citep{Taylor22,Taylor23} computed the streamlines for a sphere in a rotating flow and experimentally showed that fluid columns in solid-body motion are created parallel to the axis of rotation when the flow is perturbed. These structures are called Taylor columns and are relevant for the flow topology in centrifuges. More generic conical structures are normally observed in rotating flows and influence the motion of particles in the far field \citep{Herron75}. Other studies on particles in rotating flows are due to \cite{Mason75}, who computed the force on a particle of arbitrary shape moving transversally through a rotating fluid, \cite{Miyazaki95}, who computed the force on a sphere translating relative to a shear and a rotating flow, and showed that pure rotation increases the drag on the particle, and \cite{Hocking79}, who computed the drag on a sphere for Taylor columns of length comparable to the axial length of the rotating container. A comparison with experiments is provided by \cite{Karanfilian81}. Further investigations included determining the mobility of a sphere moving along the axis of rotation of the surrounding creeping flow \citep{Weisenborn85} and Stokes' drag and Kirchhoff's couple corrections for non-zero Reynolds and Taylor numbers. \cite{Candelier05} theoretically analyzed the motion of a spherical particle and a spherical bubble in a solid-body rotation flow, validating their analytic results with experimental measurements. They showed that the Coriolis force induces a drag and lift correction which remarkably reduce the particle migration rate away from the axis. Other important effects on the dynamics of a particle are given by the deflection of its wake, which might make the particle spinning faster in a flow with a significant cross-stream shear, taking over the inertial effects, which tend to slow down the particle rotation \citep{Bluemink08}. More recent investigations focused on the effect of inertia on drag and lift force on a particle in a rotating flow making use of theoretical analyses, experimental measurements and numerical simulations \citep{Candelier08,Bluemink10}.

A recent publication of \cite{Xu16} has shown that ordinary centrifugation typical of particles in steady vortices can be opposed by Coriolis forces if the vortex is pulsating. This gives rise to {\it oscillatory counter-centrifugation}, which attracts heavy particles to the vortex core and repells light ones away from the vortex core if the pulsation frequency of the vortex is high enough. {\color{black}This phenomenon is relevant to microfluidic systems integrated in rotating bio-disk platforms, which are used for the integration and automation of life science analysis and synthesis protocols\citep{Ducree07}. Other possible applications involve lab-on-a-chip devices, where centrifugal microfluidics is finding a growing interest \cite{Mark10}. We will study the motion of a single particle in a simple pulsating vortex, which ideally corresponds to an oscillatory rigid body rotation of the disk and of the fluid flow in a liquid-filled insert fixed on the bio-disk. Our focus will be on the motion of a single particle, and we aim at demonstrating the surprisingly complex particle dynamics in pulsating vortex flows which has been overlooked so far. Our simplified approach aims at uncovering the new aspects of such a physical mechanism, and we will therefore avoid to include confinement effects on the particle trajectory.}

The remainder of this paper {\color{black}is structured} as follows: in Sec.\ \ref{sec:Mathematics} we formulate the mathematical problem which governs the motion of a small rigid particle in a pulsating vortex flow, Sec.\ \ref{sec:Numerics} presents the numerical methods employed in our study and validates the implementation of our codes, Sec.\ \ref{sec:Results} gathers the results of our investigation, which are then summarized and discussed in Sec.\ \ref{sec:DiscussConcl}.

\section{Problem formulation}\label{sec:Mathematics}

The motion of a small rigid spherical particle of radius $a$ and density $\rho_\text{p}$ in an unbounded incompressible fluid of kinematic viscosity $\nu$ and density $\rho_\text{f}$ is modeled by the Maxey--Riley equation \citep{Maxey83}
\begin{align}\label{eq:Maxey-RileyDimensional}
\rho_\text{p} \cfrac{\dd \vec{v}}{\dd t}=& \rho_\text{f} \cfrac{\DD \vec{u}}{\DD t} + \left(\rho_\text{p} - \rho_\text{f}\right) \vec{g}  -\cfrac{9\nu\rho_\text{f}}{2 a^2}\left(\vec{v} -\vec{u} -\cfrac{a^2}{6}\nabla^2\vec{u}\right) -\cfrac{\rho_\text{f}}{2}\left[\cfrac{\dd \vec{v}}{\dd t} -\cfrac{\DD}{\DD t}\left(\vec{u} +\cfrac{a^2}{10}\nabla^2\vec{u}\right)\right]\nonumber\\
& -\cfrac{9\rho_\text{f}}{2 a}\sqrt{\cfrac{\nu}{\pi}}\int_0^t{\cfrac{1}{\sqrt{t-\tau}}\cfrac{\dd}{\dd \tau}\left(\vec{v}-\vec{u}-\cfrac{a^2}{6}\nabla^2\vec{u}\right)\dd\tau}, 
\end{align}
where $t$ denotes the time, $\vec{v}$ and $\vec{u}$ are the particle and the fluid velocity, respectively, and $\vec{g}$ is the gravity acceleration. On the left-hand side of \eqref{eq:Maxey-RileyDimensional} there is the rate of change of the particle momentum, whereas the right-hand side gathers, in order, the force exerted on the particle by the undisturbed flow, buoyancy, Stokes drag, added mass and Basset history force; all those terms which include $a^2\nabla^2\vec{u}$ represent the Fax\'en correction \citep{Faxen22}. Two different notations are employed for the material derivatives $\dd_t$ and $\DD_t$ because they respectively refer to the Lagrangian derivative along the particle trajectory 
\begin{equation}
\cfrac{\dd \vec{A}}{\dd t} = \cfrac{\partial \vec{A}}{\partial t} + \left(\vec{v}\cdot\nabla\right)\vec{A}
\end{equation}
and along the fluid trajectory
\begin{equation}
\cfrac{\DD \vec{A}}{\DD t} = \cfrac{\partial \vec{A}}{\partial t} + \left(\vec{u}\cdot\nabla\right)\vec{A},
\end{equation}
where $\vec{A}$ is a generic vector field and $\partial_t$ denotes the Eulerian derivative.

Building upon the Kirchhoff vortex, we derive a two-dimensional time periodic incompressible flow. Rescaling length, velocity and time by $L$, $U$ and $L/U$, respectively, where $L$ and $U$ are the characteristic length and velocity of the fluid flow, the stream function $\psi$ and the fluid flow velocity $\vec{u}$ read
\begin{equation}\label{eq:ModelFlow}
\psi = \cfrac{\Omega}{4} (x^2 + y^2)\sin(\omega t),\quad \vec{u} = \cfrac{\Omega}{2}\left( y , - x \right)\sin(\omega t)
\end{equation}
where $\vec{x}=(x,y)$ denotes the spatial coordinate, $\Omega=10^3$ represents the amplitude of the vorticity and $\omega$ its pulsation frequency. The model flow \eqref{eq:ModelFlow} consists of an unbounded circular vortex, which changes its sense of rotation from counterclockwise to clockwise and viceversa. The flow topology is readily characterized considering that the time-periodic Kirchhoff vortex admits only one critical point at $(x,y)=(0,0)$ and the flow is rotationally-invariant with respect to such an elliptic point. Moreover, \eqref{eq:ModelFlow} does not allow for chaotic advection because all the fluid elements are forced to move along a certain periodic pathline with constant radial coordinate $r_0=\sqrt{x_0^2+y_0^2}$ determined by the initial location $(x_0,y_0)$ of the fluid element. 

Hereinafter, the gravitational forces are neglected and the particle immersed in \eqref{eq:ModelFlow} is initially velocity-matched to the fluid flow. Hence, the dimensionless Maxey--Riley equation \citep{Farazmand15} reads
\begin{equation}\label{eq:Maxey-RileyNonDimensional}
\cfrac{\dd \vec{v}}{\dd t}= \cfrac{1}{\varrho + 1/2} \left[  \cfrac{3}{2} \cfrac{\DD \vec{u}}{\DD t} - \cfrac{\varrho}{\St} \left( \vec{v}-\vec{u} \right) -\sqrt{\cfrac{9\varrho}{2 \pi\St}}\int_0^t{\cfrac{1}{\sqrt{t-\tau}}\cfrac{\dd}{\dd \tau}\left(\vec{v}-\vec{u}\right)\dd\tau}\right], 
\end{equation}
where two non-dimensional groups arise: the particle-to-fluid density ratio $\varrho$ and the Stokes number $\St$
\begin{equation}\label{eq:NonDimensionalGroups}
\varrho=\cfrac{\rho_\text{p}}{\rho_\text{f}}\quad ,\quad \St=\cfrac{2a^2}{9L^2}\varrho\Rey \quad , \quad \Rey = \cfrac{U L}{\nu}.
\end{equation}
In \eqref{eq:NonDimensionalGroups}, the Reynolds number $\Rey$ is conventionally defined. Furthermore, we stress that the choice of the model flow \eqref{eq:ModelFlow} greatly simplifies our study since $\nabla^2\vec{u}\equiv 0$, so the Fax\'en correction in \eqref{eq:Maxey-RileyDimensional} is identically zero. 

The particle trajectory is computed discretizing \eqref{eq:Maxey-RileyNonDimensional} by using a fourth-order Runge--Kutta method for the first three terms \citep[see][and Sec. 3 for a validation of the code]{Romano17c} and employing a predictor corrector method \citep{Daitche13,Xu15} for dealing with the Basset-history term. The sphere is initialized at $(x_0,y_0)=(-10,10)$, i.e. $r_0=10\sqrt{2}$ and the Stokes number is chosen to be $\St=1$. A two-dimensional parameter space is explored, selecting $\omega\in[10,200]$ and $\varrho\in[0.1,10]$.

\section{Numerical simulations}\label{sec:Numerics}
\begin{figure}[!ht]
\begin{center}
\includegraphics[width=1.0\textwidth]{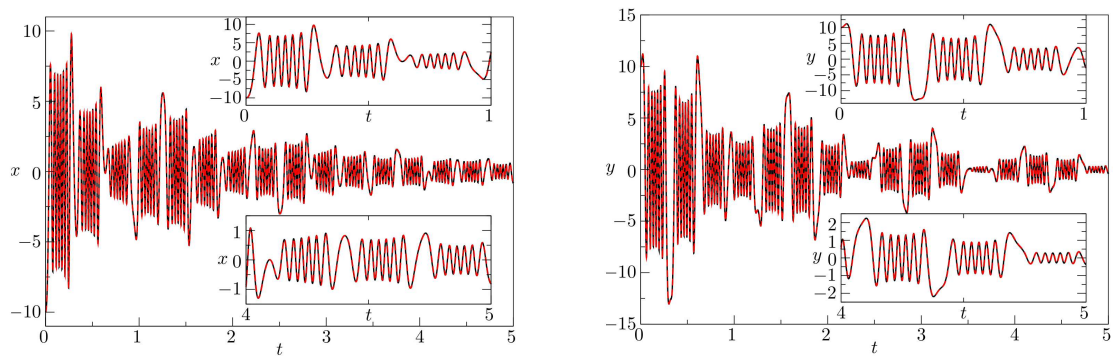}
\caption{Coordinates of the particle trajectory computed by MATLAB (black solid line) and by our 4th-order Runge--Kutta, 3/8-rule (red dashed line).}
\label{fig:RK4validation}
\end{center}
\end{figure}

The discretization of the Maxey--Riley equation without Basset-history force is carried out by means of the 4th-order Runge--Kutta, 3/8-rule. 
The implementation of our 4th-order Runge--Kutta algoritm has recently been tested in \cite{Romano17c}. A further validation is here proposed for one specific case of our investigation, testing the choice of the selected $\Delta t=10^{-3}$ in comparison with the results produced by the routine \texttt{ode45} of MATLAB setting absolute and relative error estimate to $10^{-8}$. The case selected for our benchmark employs $\Omega=10^3$, $\St=1$, $\omega=10$, $\varrho=8$ and fig.\ \ref{fig:RK4validation} depicts the particle trajectory for a particle initialized at $(x_0,y_0)=(-10,10)$ velocity-matched to the fluid flow, i.e. $\vec{v}_0=\vec{u}(t=0)$. Figure \ref{fig:RK4validation} validates our implementation of the algorithm and confirms that $\Delta t=10^{-3}$ produces results accurate enough to reliably investigate the character of the particle dynamics.

In order to discretize the full Maxey--Riley equation, including the Basset-history term, the 4th-order Runge-Kutta algorithm is supplemented by an explicit estimate of the history force. Considering that our Basset-force has the form
\begin{equation}\label{eq:BassetGeneralForm}
\vec{I}(t)=\int_{0}^{t}\cfrac{\vec{\dot{w}}(\tau)}{\sqrt{t-\tau}}\dd \tau
\end{equation}
where $\vec{w}=\vec{v}-\vec{u}$, we employ the predictor-corrector algorithm of \cite{Xu15} for estimating $\vec{I}$ at time $t=t_n$
\begin{equation}\label{eq:PredictorCorrectorMethod}
\vec{I}_n = \cfrac{2}{\Delta t} \sum_{l=2}^{n}\left(\sqrt{t_n-t_{l-1}}-\sqrt{t_n-t_{l}}\right)\left(\vec{w}_l-\vec{w}_{l-1}\right).
\end{equation}

\begin{figure}[!ht]
\begin{center}
\includegraphics[width=0.6\textwidth]{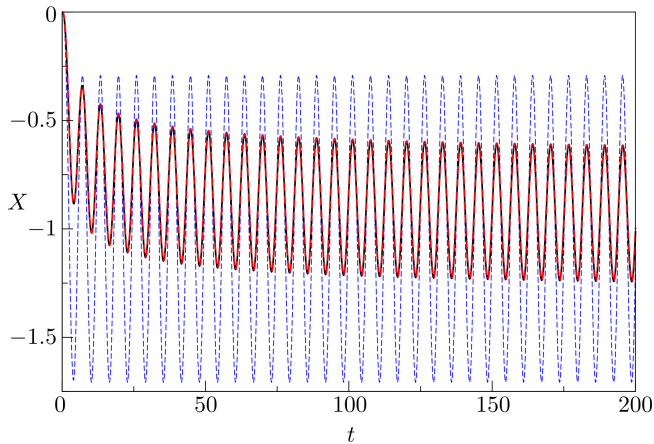}
\caption{Analytic (black solid line) and numerical solution (red dashed line) for the one-dimensional trajectory of a particle immersed in an oscillating flow. The blue dashed line denotes the numerical solution if the Basset term is neglected.}
\label{fig:Bassetvalidation}
\end{center}
\end{figure}

Our discrete approach is then tested discretizing the motion of a particle in an oscillating box, for which it yields
\begin{equation}\label{eq:OscillatoryBox}
\ddot{X}(t)=-\beta \dot{X}(t) + \left(1-\alpha\right)\sin(t) - \gamma \int_{0}^{t}\cfrac{\ddot{X}(\tau)}{\sqrt{t-\tau}}\dd\tau,
\end{equation}
where $X$ denotes the particle position, $X(t=0)=0$, $\dot{X}(t=0)=0$, and $\alpha$, $\beta$ and $\gamma$ are three parameters which characterize the particle motion. This one-dimensional problem is formally of the same class of the Maxey--Riley equation and admits an analytic solution in closed form given by \cite{Xu15}
\begin{equation}\label{eq:AnalyticalSolution}
X(t) = \left(1-\alpha\right)\left[\sum_{j=1}^{6}A_j R_j \exp \left(R_j^2 t \right)\text{Erfc}\left(-R_j\sqrt{t}\right)+A_7\right]
\end{equation}
where the coefficients reported in \eqref{eq:AnalyticalSolution} are 
\begin{align}\label{eq:AnalyticalSolutionCoeff}
R_1 &= \exp\left(\cfrac{\pi}{4}i \right),\ \  R_2 = \exp\left(\cfrac{3\pi}{4}i \right),\ \ R_3 = \exp\left(\cfrac{5\pi}{4}i \right),\nonumber \\
R_4 &= \exp\left(\cfrac{7\pi}{4}i \right), \ \ R_{5,6} = \cfrac{-\gamma\sqrt{\pi} \pm \sqrt{\gamma^2\pi-4\beta}}{2}, \nonumber \\
A_j &= \cfrac{1}{R_j^2\prod_{m=1,m\neq j}^{m=6}\left(R_j-R_m\right)} \ \ (j=1, \ 2, \ \dots , \ 6), \nonumber \\
A_7 &= \beta^{-1}.
\end{align}
Figure \ref{fig:Bassetvalidation} depicts the exact solution (black solid line) and the numerical prediction (red dashed line) for $\alpha=2$, $\beta=1$ and $\gamma=1$ in the time interval $t\in[0,200]$ employing $\Delta t = 10^{-3}$. The blue dashed line shows the numerical solution obtained neglecting the Basset-history force to make clear the importance of discretizing it. The very good agreement between our numerical prediction (red dashed line) and the exact solution (solid black line) successfully concludes the validation of our code.

\section{Results}\label{sec:Results}

A recent study has shown that the ordinary centrifugation of an heavy particle in a stationary vortex can be reverted in a pulsating vortex, leading to oscillatory counter-centrifugation \citep[OCC][]{Xu16}. The phenomenon is explained by considering the Coriolis force, which may oppose and even dominate inertial centrifugation. Making use of the method of averaging and simplifing the equation of motion for the particle, \cite{Xu16} predict that OCC always occurs if $3\sqrt{3}\nu<\omega a^2\sqrt{2\varrho+1}$. In the following we will demonstrate that the dynamics of a spherical particle in an harmonically oscillating vortex is far more complex than that, and it represents a dynamical system with several interesting features. {\color{black}Moreover, we stress that the threshold predicted by \cite{Xu16} is valid only under the assumption that $\dot{r}/r$ is negligible, which is not the case for large $\Omega$ and $\omega$. The theoretical interest towards these regimes motivates the choice of our parameters.}

\begin{figure}[!ht]
\begin{center}
\includegraphics[width=0.7\textwidth]{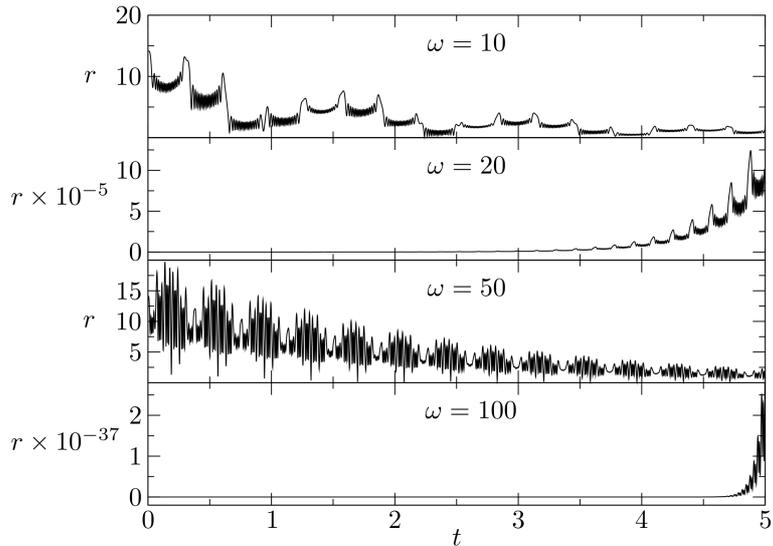}
\caption{Radial coordinate of a particle trajectory for $\Omega=10^3$, $\varrho=4$, $\St=1$ and $\omega\in\left\lbrace10, 20, 50, 100\right\rbrace$. The trajectories are integrated for $t=5$.}
\label{fig:RadialCoordinate}
\end{center}
\end{figure}

In this first part of our study, the Basset-history force is neglected. By integration of the particle trajectory for $t\in[0,5]$, we make sure that the sphere has overcome the initial transient phase and finally converged towards the attractor or diverged away from the repellor. Figure \ref{fig:RadialCoordinate} depicts the radial coordinate for four particle trajectories of a sphere with $\varrho=4$ when the vortex pulsation frequencies are $\omega=10$, $20$, $50$ and $100$. Since the particle is heavier than the fluid, it would be centrifuged away from the vortex core if the vortex were steady. As predicted by \cite{Xu16}, the oscillatory counter-centrifugation occurs already for $\omega=10$. Surprisingly, the radial coordinate of the particle trajectory $r(t)$ for $\omega=20$ shows that OCC is a reversible mechanism, in contrast to what suggested by \cite{Xu16}. The changing in character of the critical point at the vortex center is further observed passing from $\omega=20$ (the vortex core is a repellor) to $\omega=50$ (the vortex core turns into an attractor) to $\omega=100$ (the vortex core turns back into a repellor). {\color{black} The radial coordinate of the particle trajectory for $\omega=20$ and $\omega=100$ is $r\gg 1$ and it clearly shows the divergence trend due to centrifugation. We however remark that in the presence of actual confinement effects (e.g. particle suspended in a insert of a bio-disk), the particle would be settling on the outer wall of the cavity/disk, balancing centrifugal forces with the boundary repulsion due to presence of the cavity walls.}

\begin{figure}[!ht]
\begin{center}
\includegraphics[width=1.0\textwidth]{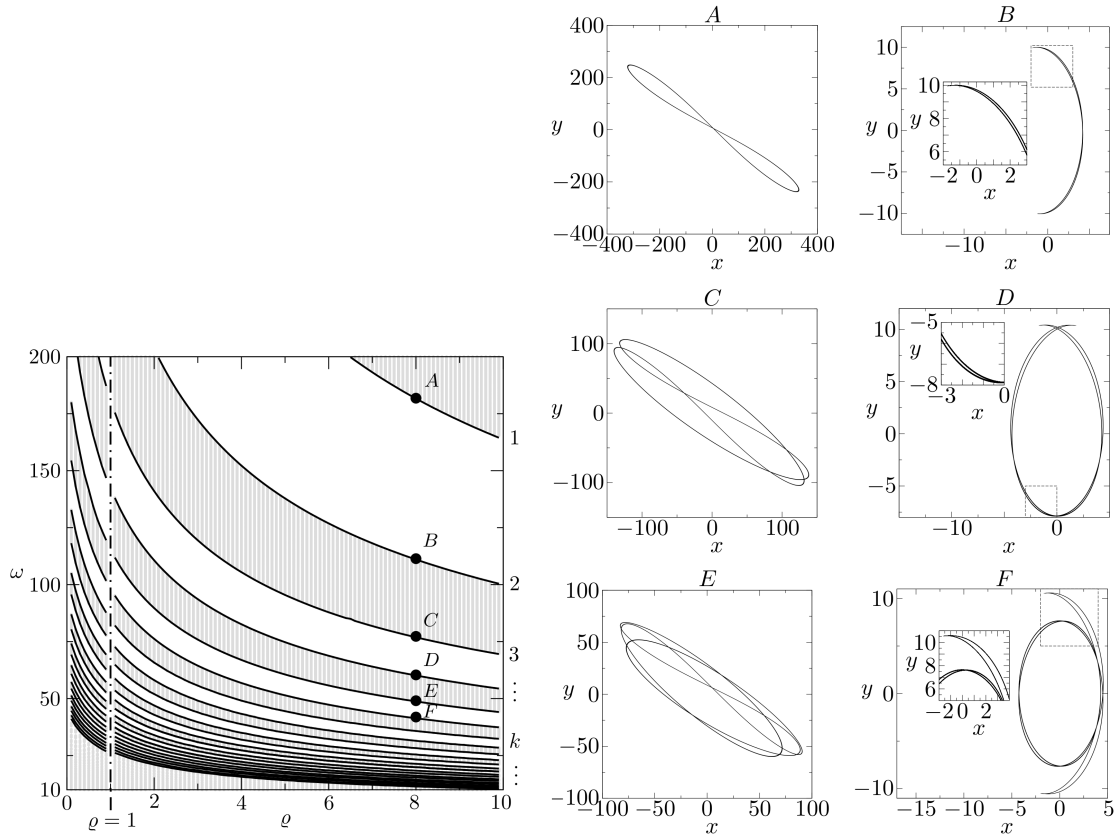}
\caption{Left: Parameter space for $\St=1$, $\Omega=10^3$ spanned by $\omega\in[10,200]$ and $\varrho\in[0.1,10]$ in which the converging (grey region) and diverging (white region) particle trajectories are identified, together with six of the non-trivial particle attractors (solid lines) which define the boundaries between centrifugal and centripetal motion of the particles. Right, A--F: Particle attractors for $\varrho=8$.}
\label{fig:ParameterSpace}
\end{center}
\end{figure}

The investigation of the parameter space is conducted screening the frequency range with $\Delta\omega=0.1$ and the density-ratio range with $\Delta\varrho=0.1$, for a total of {\color{black}approximately} $2\times 10^5$ simulations. The outcome of such numerical study is depicted in fig.\ \ref{fig:ParameterSpace}. The grey regions denote parameters for which the particle is asymptotically attracted towards the vortex core, whereas the white areas indicate an asymptotic centrifugal motion for the particle. The dashed-dotted line refers to $\varrho=1$, for which a particle is neither attracted nor repelled by the vortex core and represents a critical line for the parameter space, since light particles ($\varrho<1$) behave opposite of heavy particles ($\varrho>1$). The solid lines, progressively labeled on the right side of the left panel, denote couples $(\omega,\varrho)$ for which centrifugal and centripetal forces are balanced in average and non-trivial limit cycles are formed. The shape of the attractors along each of these curves is qualitatively identical, while switching from a curve to another, the particle limit cycle experience a qualitative change. Six examples are reported in the right panels of fig.\ \ref{fig:ParameterSpace} for $\varrho=8$ (labels A--F), for which very intricate  cat's eye, period-2 and period-4 patterns are observed. The creation of such new attractors for the particle trajectory goes far beyond the notion of oscillatory counter-centrifugation introduced by \cite{Xu16} and helps to explain the complex dynamics observed in fig.\ \ref{fig:RadialCoordinate}. Les us consider, for instance, the case $\omega=10$ for which four different time scales are observed. The longest (asymptotic, $t\rightarrow \infty$) and the shortest ($t=\mathcal{O}(0.01)$) time scale are common to all the cases depicted in fig.\ \ref{fig:RadialCoordinate}. However, for $\omega=10$, a third time scale is identified between two local spikes ($t=\mathcal{O}(0.3)$) and a forth time scale characterizes the pattern formed by a batch of spikes ($t=\mathcal{O}(1)$). This peculiar centripetal trend can be understood considering that the particle phase space for $\Omega=10^3$, $\St=1$, $\varrho=4$ and $\omega=10$ is influenced by the limit cycles which are going to be created between $\omega=10$ and $\omega=20$. The qualitative different patters observed for $\omega\in\left\lbrace10, 20, 50, 100\right\rbrace$ can then be understood considering that the limit cycles leaving their footprint on the particle dynamics are qualitatively very different (see fig.\ \ref{fig:ParameterSpace}(A--F)) since they belong to different boundary lines between centripetal and centrifugal asymptotic motion (see solid lines in fig.\ \ref{fig:ParameterSpace}). Proposing a nomenclature different from oscillatory counter-centrifugation seems therefore appropriate, in order to highlight the complexity and reversibility of our phenomenon. Hence, we will rather term it {\it oscillatory switching centrifugation} (OSC), with the boundaries between centrifugal and centripetal motion being the {\it centrifugal switching boundaries} (CSB, black lines in fig.\ \ref{fig:ParameterSpace}).

\begin{figure}[!ht]
\begin{center}
\includegraphics[width=0.5\textwidth]{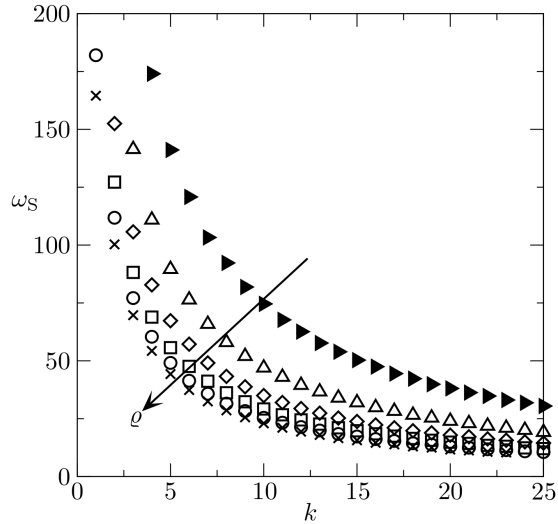}
\caption{Switching frequencies $\omega_\text{S}$ along the first 25 CSB (enumerated by $k$) for $\Omega=10^3$, $\St=1$ and $\varrho=10$ ($\times$), $8$ ($\circ$), $6$ ($\square$), $4$ ($\diamond$), $2$ ($\bigtriangleup$) and $0.5$ ($\blacktriangleright$).}
\label{fig:SelfSimilarity}
\end{center}
\end{figure}

An interesting feature of OSC highlighted by fig.\ \ref{fig:ParameterSpace} is the self-similarity of the centrifugal switching boundaries. The first 25 CSB are identified and shown in fig.\ \ref{fig:ParameterSpace} as solid lines, and they all seem to belong to the same family of curves. Another factor in support of their self-similarity is reported in fig.\ \ref{fig:SelfSimilarity}, where the switching frequencies $\omega_\text{S}$ along these 25 CSB (enumerated by the index $k$) are reported for six density ratios: $\varrho=10$, $8$, $6$, $4$, $2$ and $0.5$. 

\begin{figure}[!ht]
\begin{center}
\includegraphics[width=0.5\textwidth]{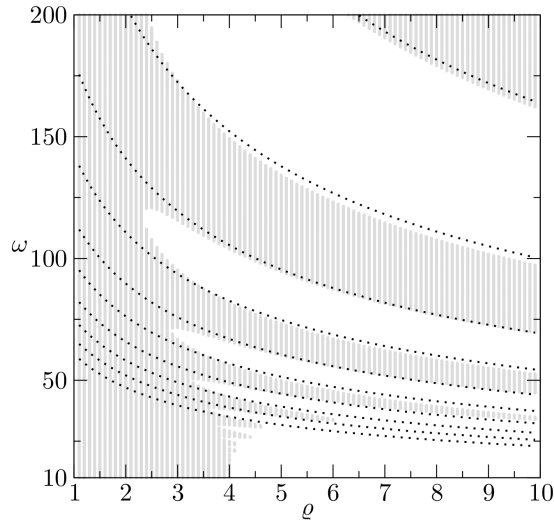}
\caption{Parameter space for $\St=1$, $\Omega=10^3$ spanned by $\omega\in[10,200]$ and $\varrho\in(1,10]$ in which the converging (grey region) and diverging (white region) particle trajectories are identified, including the effect of the Basset-history force. The dotted lines denote the first ten CSB computed neglecting the history force.}
\label{fig:Basset}
\end{center}
\end{figure}

All the computations reported so far are obtained neglecting the Basset-history force. Figure \ref{fig:Basset} shows how the parameter space changes for $\St=1$, $\Omega=10^3$, $\omega\in[10,200]$ and $\varrho\in(1,10]$ when the history force is included in the particle trajectory. Once again, asymptotically speaking, the grey regions denote centripetal motion, while the white areas refer to centrifugal motion. A direct comparison between fig.\ \ref{fig:ParameterSpace} and fig.\ \ref{fig:Basset} is offered by the dotted lines, which depict the first ten CSB ($k\in[1,10]$) of fig.\ \ref{fig:ParameterSpace}. The Basset-history force has a relevant impact on the asymptotic trend of the particle trajectory only for particles slightly density-mismatched and for relatively low oscillation frequencies. For $\varrho>4$ and $\omega>30$ the differences between fig.\ \ref{fig:ParameterSpace} and fig.\ \ref{fig:Basset} can hardly be observed.

\section{Discussion and Conclusion}\label{sec:DiscussConcl}

The motion of a small rigid spherical particle in a pulsating vortex {\color{black}has been studied} by means of numerical simulations based on the Maxey--Riley equation. Due to Coriolis forces, the ordinary centrifugation observed in steady vortices can be reverted if the vortex is pulsating at high frequency, leading to a centripetal motion for heavy particles and a centrifugal motion for light particles. This phenomenon has been called {\it oscillatory-counter centrifugation} by \cite{Xu16}. 

In our study we discover that changing the pulsation frequency of the vortex, even the OCC can be reverted, giving rise to regions, in parameter space, where ordinary centrifugation occurs, and other regions where counter-centrifugation occurs. We termed this phenomenon {\it oscillatory switching centrifugation} (OSC) and demonstrated it for the first time in all its complexity, making use of the simplest model flow. 

Despite the simple background flow employed, we speculate OSC has a remarkable relevance in far more complex systems. Let us consider, for instance, the segregation of particles heavier than the fluid. Making use of ordinary finite centrifuges, it is not possible to asymptotically separate heavy particles with different densities. On the other hand, based on fig.\ \ref{fig:Basset}, one can find a frequency, for instance $\omega=100$, which  attracts $(\varrho=4)$-particles to the vortex core and centrifuges out $(\varrho=6)$-particles. {\color{black} Another application of our study considers the protocols employed by biologists in microcentrifuges for breaking down the membrane of a cell (lysis). Other relevant applications involve lab-on-a-chip devices, where microfluidic centrifugation is frequently involved.}

Finally, because of the following interesting features, we expect OSC to have an impact on theoretical and experimental studies on particle dynamics: (i) the Hamiltonian system represented by the two-dimensional fluid element dynamics \eqref{eq:ModelFlow} is described by a relatively simple model which can be experimentally realized in a controlled environment, (ii) the absence of chaotic fluid pathlines simplifies the theoretical analysis, (iii) the dynamics of particles immersed in this flow show self-similar characters whose full understanding would benefit from a renormalization theory analysis and (iv) several families of non-trivial particle limit cycles occur, whose footprint characterizes the particle dynamics. Hence, we speculate that OSC offers an interesting case of study for the dynamical system community and a novel framework for challenging \citep{Xu16} experimental investigations.

\bibliographystyle{jfm}
\bibliography{Bibliography}

\end{document}